\documentclass{elsart}
\newcommand{\lesssim}{\raisebox{0.3mm}{\em $\, <$} 
\hspace{-3.3mm} \raisebox{-1.8mm}{\em $\sim \,$}}
\newcommand{\gtrsim}{\raisebox{0.3mm}{\em $\, >$} 
\hspace{-3.3mm} \raisebox{-1.8mm}{\em $\sim \,$}}
\begin{document}

\begin{frontmatter}

\title{Summary of Working Group 1: Theory Part}

\author{Osamu Yasuda}
\ead{yasuda@phys.metro-u.ac.jp}

\address{Department of Physics, Tokyo Metropolitan University \\
 1-1 Minami-Osawa Hachioji, Tokyo 192-0397, \cny Japan}

\begin{abstract}
I will summarize theoretical issues in Working Group 1 at Nufact'01.
Although there may not be complete agreement yet on the exact optimum
baseline $L$ and the muon energy $E_\mu$ for measurements of the CP
phase at a neutrino factory, all the works done so far indicate that
the optimum set ($L$, $E_\mu$) tends to be smaller than (3000km,
50GeV) if the uncertainty of the matter effect is assumed to be
larger than $\pm$5\% or the background fraction is much larger than
$10^{-5}$.
\end{abstract}

\end{frontmatter}

\section{Model Building}
In the three flavor framework of neutrino oscillations,
the only two parameters whose lower bound is not known
in the mixing matrix $U_{\rm MNSP}$ \footnote{
Petcov \cite{petcov} proposed that the neutrino
mixing matrix be referred to as the Pontecorvo-Maki-Nakagawa-Sakata
\cite{Maki:1962mu,Pontecorvo:1968fh}
matrix.  Here I refer to the matrix as $U_{\rm MNSP}$.}
are $\theta_{13}$ and $\delta$, and the magnitude of
$|U_{e3}|=s_{13}$ is expected to give us a clue
to model buildings.
Tanimoto \cite{Tanimoto:2001aq} gave various examples of
models which predict the magnitude of $U_{e3}$.
Some (e.g., anarchy models) predict large
$U_{e3}\sim O(\lambda)$ where $\lambda\simeq 0.2$,
other models (those with some flavor symmetry or with the conformal
field fixed point) predict $U_{e3}\sim O(\lambda^2-\lambda)$
or $O(\lambda^3-\lambda)$, while Zee type models typically
give tiny $U_{e3}\ll\lambda^3$.  Some GUT models predict testable
sum rules among $V_{\rm CKM}$ and $U_{\rm MNSP}$.
Koide \cite{Koide:2001dr} emphasized that the deviation of
the solar mixing angle
from the maximal mixing is small in the Zee model:
$1-\sin^22\theta_\odot\le(1/16)
(\Delta m_\odot^2/\Delta m_{\mbox{\rm\scriptsize atm}}^2)^2.$
This relation can be checked experimentally
in the near future.
Kitabayashi \cite{Kitabayashi:2001nw} presented
a model with a minimal $SU(3)_L\times U(1)_N$ gauge symmetry,
an approximate $L_e-L_\mu-L_\tau$ symmetry and a discrete
$Z_4$ symmetry.  This model yields the LMA MSW solution
and explains the hierarchy
$|\Delta m_\odot^2|\ll|\Delta m_{\mbox{\rm\scriptsize atm}}^2|$
by loop corrections.
Leung \cite{Leung:2001zc} considered effective operators relevant for
generating small Majorana neutrino masses.  All the effective higher
dimensional ($5\le d \le 11$) operators are compiled which conserve
baryon number and the SM gauge invariance but violate lepton number by
two units.  If neutrinoless double $\beta$ decays are observed, then a
number of operators will be reduced.
Geer \cite{geer} discussed how precise measurements of the oscillation
parameters would give us useful information at high energy scale, by
taking a specific SUSY SO(10) GUT model with the LMA MSW solution.  In
this particular example, superbeam experiments and/or neutrino
factories will constrain the possible values of the GUT-scale
parameters.

\section{Analysis of solar and atmospheric neutrinos and $(\beta\beta)_{0\nu}$}

Lisi \cite{lisi} reported on analysis of the atmospheric neutrino
data in two, three and four flavor frameworks.
In the two flavor framework it was shown that 
$\nu_\mu\leftrightarrow\nu_\tau$ oscillation with maximal mixing
is robust.  Also in the three flavor scenario,
by combining with the CHOOZ data the dominant oscillation
in the atmospheric neutrino was shown to be
$\nu_\mu\leftrightarrow\nu_\tau$, possibly with small
contribution of $\nu_\mu\leftrightarrow\nu_e$.
It was pointed out that non-hierarchical scenario where
$\Delta m^2_\odot\sim\Delta m^2_{\mbox{\rm\scriptsize atm}}$ holds are
not totally excluded yet.
In the four flavor case the data were analyzed with
the so-called (2+2)-scheme, and some contribution of
$\nu_\mu\leftrightarrow\nu_s$ is still allowed at present.

Pe\~na-Garay \cite{penya} presented analysis of the solar
neutrino data in two, three and four flavor mixings.
In the case of two flavor active oscillations,
the LMA MSW solution is the best fit, and it is followed by
the LOW solution.  The SMA MSW and vacuum solutions seem to
be disfavored now.  In the two flavor sterile oscillations,
the SMA MSW solution is the best fit.
In the three flavor scenario, with the CHOOZ constraint,
the dominant channel is
$\nu_e\leftrightarrow (\nu_\mu-\nu_\tau)/\sqrt{2}$,
i.e., $\nu_e\leftrightarrow \nu_\mu$ and
$\nu_e\leftrightarrow \nu_\tau$ oscillations occur
with equal weight.
In the four neutrino mixing with the (2+2)-scheme,
the oscillation probability is characterized by
one additional parameter $|U_{s1}|^2+|U_{s2}|^2$
which becomes 0 (1) for pure active (sterile) oscillation,
respectively.  By combining the analyses of the solar and atmospheric
neutrino data, it was found before the SNO data
that the case with pure $\nu_e\leftrightarrow \nu_s$
in the solar neutrinos and pure $\nu_\mu\leftrightarrow \nu_\tau$
in the atmospheric neutrinos is close to the best fit.
Even after the SNO data this solution is not excluded
yet \cite{Barger:2001zs,Gonzalez-Garcia:2001zi}.

Petcov \cite{petcov} gave a talk on the neutrino mass spectrum
and CP violation
in the lepton sector in the three flavor framework.
Neutrinoless double $\beta$ decay experiments in
the future can provide information on the lightest neutrino mass and
the CP violation in the lepton sector.  In particular, if a positive
result ($m_{\nu_e}\ge 0.35$eV) is reported from
neutrinoless double $\beta$ decay experiments, then
combining the data of ${}^3$H $\beta$ decay experiments
of the KATRIN project, it would allow us to determine
$m_1, m_2, m_3$ and to know the existence of the CP violation
in the lepton sector.

\section{Various issues in neutrino factory and superbeams}

In the past there have been a lot of discussions on the physics
potential of neutrino factories and conventional superbeams, and
people investigated the optimum baseline and the muon
(or neutrino) energy to
determine $\theta_{13}$, the sign of $\Delta m^2_{32}$ and the CP
phase $\delta$ of the MNSP matrix in the three flavor framework.
Recent progress includes the effects of correlations of errors of the
oscillation parameters, the background effects, etc.  The ultimate
purpose of neutrino factories and conventional superbeams is to
determine $\delta$.  So far all the discussions are based on indirect
measurements of CP violation, i.e., assuming
the three flavor mixing,
a quantity
\begin{eqnarray}
\Delta \chi^2_{\mbox{\rm{\scriptsize indirect}}}= \sum_j {\left[
N_j(\delta)-N_j(\delta=0)\right]^2 \over \sigma^2_j},
\label{eqn:indirect}
\end{eqnarray}
which compares
the difference of the cases of $\delta\ne0$ and $\delta=0$
with the errors, is optimized.
In (\ref{eqn:indirect}) $N_j(\delta)$ stands for a binned number of events with
$\delta$ and it is understood that
$\Delta \chi^2_{\mbox{\rm{\scriptsize indirect}}}$ is optimized
with respect to other parameters in $N_j(\delta=0)$.  
As has been criticized by some people \cite{Sato:2000wv,konaka}, this
approach does not directly deal with CP violating processes, since
$\nu$ and $\bar{\nu}$ are not directly compared and what we wish to
have is an analogue of $N(K_L\rightarrow 2\pi)/N(K_L\rightarrow 3\pi)$
in the $K^0-\bar{K}^0$ system \cite{Minakata:2001rj}.  However, such quantity
has not been discovered yet in the case of neutrino factories or any
very long baseline experiments, since the dependence of
the probabilities for $\nu$ and
$\bar{\nu}$ on $\delta$ and $A$ is such that
$P(\nu_\mu\rightarrow\nu_e)=f(E,L;\theta_{ij},\Delta
m^2_{ij},\delta;A)$ and $P(\bar{\nu}_\mu\rightarrow\bar{\nu}_e)
=f(E,L;\theta_{ij},\Delta m^2_{ij},-\delta;-A)$, where $f$ is a
certain function and $A\equiv \sqrt{2}G_F N_e$ stands for the matter
effect, and the quantity obtained from experiments on $\bar{\nu}$ is
$f(\cdots,-\delta;-A)$, so that comparison between $f(\cdots,\delta;A)$ and
$f(\cdots,-\delta;A)$ is impossible in a strict sense.
Therefore all the discussions
below basically use indirect measurements of CP violation.

One of the important issues in the discussions on
uncertainties of the parameters is that of the density of
the Earth.
Geller, a geophysicist working on the three dimensional structure
of the Earth, explained the density distribution in the
Earth's Interior to physicists \cite{geller}.
To determine the density, there are complications.  The neutrino path goes
through the upper mantle as well as the crust, so one
needs both crust models and regional upper mantle models
which are not well established with great accuracy
at present.
The distribution of the density is much less well
constrained than seismic velocities which are known
relatively well.  Despite all these problems, from
ballpark guesses rather than rigorous error estimation procedures,
he concluded that the accuracy of estimates
of the average density along the neutrino beam is
at worst $\pm$10\% and is probably within $\pm$5\%.
There were two talks in which a method was presented to
treat the density profile which is in general not
constant.
Takasugi \cite{takasugi} discussed T violation
and derived a formula up to third order
in $\Delta m^2_{21}L/2E$ and $\delta a(x)L/2E$ where $\delta a(x)$
stands for deviation from the PREM.  He concluded that
the asymmetric matter contribution to T violation
is negligible.
Ota \cite{ota} presented a method of
Fourier expansion of the density profile to discuss the matter effect
and its ambiguity, and showed that only the first few
coefficients are important.

Lindner \cite{lindner} discussed the optimization of sensitivity
to $(\Delta m^2_{32}, \sin^22\theta_{23})$,
$\sin^22\theta_{13}$, $(\Delta m^2_{21}, \sin^22\theta_{12})$,
and $\delta$, taking into account
the correlations of errors of
the oscillations parameters and assuming that background
effects are negligible.
In the case of $(\Delta m^2_{32}, \sin^22\theta_{23})$,
the optimum baseline lies between
1000km and 5000km.
In the case of $\sin^22\theta_{13}$,
1000km$\lesssim L\lesssim$5000km gives similar results
for $\sin^22\theta_{13}\gtrsim 10^{-2}$ and
the optimum baseline is between 7000km and 8000km
for $\sin^22\theta_{13}\lesssim 10^{-3}$.
Sensitivity to $(\Delta m^2_{21}, \sin^22\theta_{12})$
turns out to be much worse than that of the KamLAND
experiment \cite{ishihara}.
As for $\delta$, the optimum baseline is around
3000km and high energy ($E_\mu\gtrsim$40GeV) is preferred.
This result almost agrees with \cite{Cervera:2000kp} and
\cite{Pinney:2001bj}.\footnote{Strictly speaking,
there may not be exact agreement among these works
since the reference values for the oscillations parameters
($\Delta m^2_{jk}$, $\theta_{jk}$), the conditions on
the backgrounds or on the uncertainty of the matter effect
and the statistical treatments are different among these
works.  It should be clarified in the future work which
condition is important to obtain the different optimum
set ($L$, $E_\mu$).  I would like to thank P. Huber for
communications on this point.}

Burguet-Castell and Mena \cite{Burguet-Castell:2001uw}
refined their analysis in \cite{Cervera:2000kp}
of the sensitivity to $\delta$ and
$\theta_{13}$ by considering the full range of these
parameters.  They found that there is twofold degeneracy
in ($\theta_{13}$, $\delta$).  This is because
the expansion of the oscillation probabilities
to second order in $|\theta_{13}|\ll 1$ gives
a quadratic equation in $\theta_{13}$:
\begin{eqnarray}
P_{\nu_e\nu_\mu(\bar{\mu}_e\bar{\nu}_\mu)}
=X_\pm\theta_{13}^2+Y_\pm\theta_{13}\cos(\delta-{\Delta_{13}L \over 2})
+P^{sol},\nonumber
\end{eqnarray}
where notations are given in \cite{Burguet-Castell:2001ez}.
To resolve this degeneracy, a combination of
two baselines seems necessary.

Pinney \cite{Pinney:2001bj} also discussed the optimum baseline and the
muon energy of a neutrino factory for $\delta$,
by taking into account the systematic errors and
the correlations of errors of the oscillation
parameters as well as the density of the Earth.
If the background fraction $f_B$ is $10^{-5}$ and
the uncertainty of the matter effect is $\pm$5\%, then
$E_\mu\sim$50GeV and $L\sim$3000km is the optimum
parameter set (this result almost agrees with those in
\cite{Cervera:2000kp} and \cite{lindner}), but if $f_B=10^{-3}$ or
if the uncertainty of the matter effect is $\pm$10\%, then
the lower muon energy ($E_\mu\lesssim$20GeV) and
the shorter baseline ($L\sim$1000km--3000km) gives the optimum.
In \cite{Pinney:2001xw} it was shown that
$\Delta \chi^2_{\mbox{\rm{\scriptsize indirect}}}$
used in the analysis in \cite{Pinney:2001xw} has behavior
\begin{eqnarray}
\Delta \chi^2_{\mbox{\rm{\scriptsize indirect}}}
\sim \left({J \over \sin\delta}\right)^2
{1 \over E_\mu}\left(
\sin\delta+\mbox{\rm const}{\Delta m^2_{32}L \over E_\mu}
\cos\delta\right)^2
\label{eqn:largeemu}
\end{eqnarray}
for large $E_\mu$ ($J$ stands for the Jarlskog parameter),
so that sensitivity to CP violation is lost
in the large $E_\mu$ limit.
(\ref{eqn:largeemu}) also suggests that neutrino factories
with $E_\mu\sim$50GeV and $L\sim$3000km are mainly
sensitive to $\sin\delta$ instead of $\cos\delta$.\footnote{
\cite{Koike:2000jf} criticized the analysis with
$\Delta \chi^2_{\mbox{\rm{\scriptsize indirect}}}$
($=\chi^2_1(\delta_0)$ in the notation of \cite{Koike:2000jf})
by saying that $\chi^2_1(\delta_0)$ has the behavior
$\chi^2_1(\delta_0)\sim E_\mu(J/\sin\delta)^2(\cos\delta\pm1)^2$
in the large $E_\mu$ limit
and has sensitivity mainly to $\cos\delta$.
However, this argument does not include the effects due to
the correlations of errors, and once optimization with respect
to other parameters is done,
the correct behavior (\ref{eqn:largeemu}) is obtained.}

Koike \cite{koike} presented their discussions in \cite{Koike:2000jf}
on the optimum baseline and the muon energy at neutrino factories.
His main strategy is to use
\begin{eqnarray}
\Delta {\widetilde\chi}^2_{\mbox{\rm{\scriptsize indirect}}}= \sum_j {\left[
N_j(\delta)-\bar{N}_j(\delta)-N_j(\delta=0)+\bar{N}_j(\delta=0)
\right]^2 \over \sigma^2_j},
\label{eqn:chi2}
\end{eqnarray}
where $\bar{N}_j(\delta)$ stands for a suitably normalized
binned number of events
for $\bar{\nu}_e\rightarrow\bar{\nu}_\mu$ oscillations
and again optimization with respect to other parameters
in $N_j(\delta=0)$ and $\bar{N}_j(\delta=0)$ is understood.
He claimed that
$\Delta {\widetilde\chi}^2_{\mbox{\rm{\scriptsize indirect}}}$
($=\chi^2_2(\delta_0)$ in the notation of \cite{Koike:2000jf})
is sensitive to $\sin\delta$ whereas
$\Delta \chi^2_{\mbox{\rm{\scriptsize indirect}}}$
in (\ref{eqn:indirect}) is mainly to $\cos\delta$.
In my opinion, however, 
$\Delta \chi^2_{\mbox{\rm{\scriptsize indirect}}}$
in (\ref{eqn:indirect}) and
$\Delta  {\widetilde\chi}^2_{\mbox{\rm{\scriptsize indirect}}}$
in (\ref{eqn:chi2}) are basically the same, as
their large $E_\mu$ behaviors are the same after
the correlations of errors are taken into account
and both use indirect measurements of CP violation.
In fact, the conclusion by Koike that
$E_\mu\sim$10GeV and $L\sim$1000km optimizes the sensitivity
to $\delta$ almost agrees with that in \cite{Pinney:2001xw}
where the case of $\Delta C\equiv$ the uncertainty of the matter
effect=$\pm$10\%
and $f_B=10^{-5}$ was also considered (See Fig. 8 in \cite{Pinney:2001xw}).
Notice that Koike assumed that all the uncertainty
of the oscillation parameters as well as the matter effect
is $\pm$10\%.

Wang \cite{wang} also discussed the optimum baseline for
experiments with conventional superbeams and for neutrino
factories by introducing a figure of merit which is defined as
\begin{eqnarray}
F_M={P \over \sqrt{(P+f_B)/N+(gP)^2+(rP)^2}},\nonumber
\end{eqnarray}
where $P$ is the oscillation probability, $N$ is the number of events,
$f_B$ is the background fraction, $r$ is the systematic uncertainty,
and $g$ is the uncertainty of the flux and the cross sections.
He concluded that $L\simeq$2000km--3000km is preferred for
measurements of
$\sin^22\theta_{13}$, $\delta$ and the sign of $\Delta m^2_{32}$.

Minakata \cite{Minakata:2001rj} discussed the physics potential of
the phase II JHF experiment which is expected to have
4MW power and 1Mt water Cherenkov detectors.
He examined
if it is possible to determine the $\sin^22\theta_{13}$,
the sign of $\Delta m^2_{32}$ and $\delta$
by introducing a CP trajectory diagram,
in which a point sweeps out an
ellipse in the $P(\nu_\mu\rightarrow\nu_e)$ --
$P(\bar{\nu}_\mu\rightarrow\bar{\nu}_e)$ plane
as $\delta$ ranges from 0 to $2\pi$.
The behavior of the ellipse varies depending on
the sign of $\Delta m^2_{32}$, and in certain cases
it is possible to determine $\delta$ and $\Delta m^2_{32}$
in the phase II JHF experiment.  On the other hand,
if the oscillation parameters lie in unlucky regions,
then there is a twofold ambiguity in ($\delta$, $\Delta m^2_{32}$).
This ambiguity is resolved by putting another detector
at $L$=700km.

Okamura \cite{okamura} discussed the physics potential of a very long
baseline experiment ($L$=2100km) with conventional superbeams
(3GeV$\lesssim E_{\mbox{\rm\scriptsize proton}}\lesssim$6GeV).
Taking into account the
correlations of errors and the background effects, he examined
sensitivity to the sign of $\Delta m^2_{32}$, $\sin^22\theta_{13}$,
($\Delta m^2_{21}$, $\sin^22\theta_{12}$), and $\delta$.  With such a
long baseline, sensitivity to the sign of $\Delta m^2_{32}$ is very
good, but sensitivity to ($\Delta m^2_{21}$, $\sin^22\theta_{12}$) is
poor as is naively expected.  Sensitivity to $\delta$ is good for
$\sin^22\theta_{13}\gtrsim 0.05$, assuming the detector size
100kt$\cdot$yr.

\section{Four neutrino scenarios}

To account for the solar and atmospheric neutrino data
as well as the LSND result in terms of neutrino oscillations,
one has to have at least four neutrino mass eigenstates
since one would need at least three independent
mass squared differences.  Four neutrino scenarios
are classified into two categories, the (2+2)-scheme and
the (3+1)-scheme, depending on the number of degenerate
mass eigenstates separated by
$\Delta m^2_{\mbox{\rm{\scriptsize LSND}}}$.

Peres \cite{peres} reviewed the phenomenology of the (3+1)-scheme,
which is excluded by the LSND data
at 95\%CL but allowed at 99\%CL.
The upper left $3\times 3$ components
of the MNSP matrix for the (3+1)-scheme are supposed to
be close to those for the three flavor case, so it gives a good fit
to the data of the solar and atmospheric neutrino data.
The (3+1)-scheme is allowed only for
$\Delta m^2_{\mbox{\rm{\scriptsize LSND}}}\simeq$0.9, 1.7, 6.0 eV$^2$,
so the upper limit on the absolute neutrino masses would
constrain the allowed region of this scheme.
$\beta$ decay experiments, neutrinoless double $\beta$ decay experiments
and supernova neutrinos give some bounds.

Bell \cite{Bell:2001yv} discussed how relic neutrino asymmetries may
be generated in the early universe via active-sterile oscillations.
In the two flavor framework it is known that asymmetry between $\nu$
and $\bar{\nu}$ is generated through active-sterile oscillations under
a certain condition for neutrino masses and the mixing parameters.
The implications of this mechanism to Big Bang Nucleosynthesis
was given in the case of a particular (2+2)-scheme model
where a $\nu_\mu - \nu_\tau$ pair is separated by a
$\nu_e - \nu_s$ pair, and the conclusion was that
$\delta N_\nu^{\rm eff}\simeq -0.3~(+0.1)$ for positive
(negative) asymmetry, respectively.

Donini \cite{Donini:2001qv} compared the physics reach of
a neutrino factory in the (3+1)- and (2+2)-schemes.
In both schemes huge CP violating effects can be observed with
a 1Kt detector, $2\times 10^{-20}\mu$'s and $L=10-100$km in the
$\nu_\mu \rightarrow \nu_\tau$ channel.
In case a conclusive confirmation of the LSND
result is absent, it is difficult to
discriminate the (3+1)-scheme from the three flavor case
unless the mixing angle is large and it is easy to
discriminate the (2+2)-scheme from the three flavor case.

\end{document}